%% file: fido_nim.tex
\documentclass[5p,twocolumn,preprint,longtitle]{elsarticle}
\usepackage[pagewise]{lineno}
\usepackage{amsmath,amsfonts,amssymb,amsthm}
\usepackage{xspace}

\newcommand{\smalllength}{l}

\newcommand{\fig}{Fig.~}
\newcommand{\sect}{Section\xspace}

\newcommand{\ibd}{inverse beta decay\xspace}
\newcommand{\ov}{OV\xspace}
\newcommand{\id}{ID\xspace}
\newcommand{\iv}{IV\xspace}
\newcommand{\targ}{NT\xspace}
\newcommand{\gc}{GC\xspace}

\newcommand{\Pmts}{PMTs\xspace}
\newcommand{\pmt}{PMT\xspace}
\newcommand{\pmts}{PMTs\xspace}
\newcommand{\buf}{Buffer\xspace}
\newcommand{\ovinf}{OV-in\-clu\-sive fit\xspace}
\newcommand{\ovexf}{OV-ig\-nored fit\xspace}
\newcommand{\dc}{Double Chooz\xspace}
\newcommand{\adc}{ADC\xspace}
\newcommand{\xyo}{$(x,y)$ coordinates\xspace}
\newcommand{\ovt}{OV track\xspace}
\newcommand{\ovts}{OV tracks\xspace}

\newcommand{\minuit}{{\sc minuit}\xspace}
\newcommand{\migrad}{{\sc migrad}\xspace}
\newcommand{\cher}{Cher\-en\-kov\xspace}
\newcommand{\scindom}{scin\-til\-la\-tion-dom\-i\-nated\xspace}
\newcommand{\cherdom}{\cher-dom\-i\-nated\xspace}
\newcommand{\upgo}{up\-wards-going\xspace}
\newcommand{\thrugo}{through-going\xspace}
\newcommand{\Thrugo}{Through-going\xspace}
\newcommand{\downgo}{down\-wards-going\xspace}
\newcommand{\linine}{${}^9$Li\xspace}
\newcommand{\btw}{${}^{12}$B\xspace}
\newcommand{\hee}{${}^8$He\xspace}
\newcommand{\dedx}{$\mathrm{d}E/\mathrm{d}x$\xspace}

\journal{Nuclear Instrumentation and Methods A}

\begin{document}

\begin{frontmatter}

\title{Precision Muon Reconstruction in Double Chooz}

\input{authors.tex}

\begin{abstract}
We describe a muon track reconstruction algorithm for the reactor
anti-neutrino experiment \dc. The \dc detector consists of two optically
isolated volumes of liquid scintillator viewed by \pmts, and an Outer
Veto above these made of crossed scintillator strips. Muons are
reconstructed by their Outer Veto hit positions along with timing
information from the other two detector volumes. All muons are fit
under the hypothesis that they are \thrugo and ultrarelativistic. If
the energy depositions suggest that the muon may have stopped, the
reconstruction fits also for this hypothesis and chooses between the two
via the relative goodness-of-fit. In the ideal case of a \thrugo muon
intersecting the center of the detector, the resolution is $\sim$40\,mm
in each transverse dimension. High quality muon reconstruction is
an important tool for reducing the impact of the cosmogenic isotope
background in \dc.
\end{abstract}


\end{frontmatter}

\section{Introduction} 

\begin{figure}

\begin{center}

\includegraphics{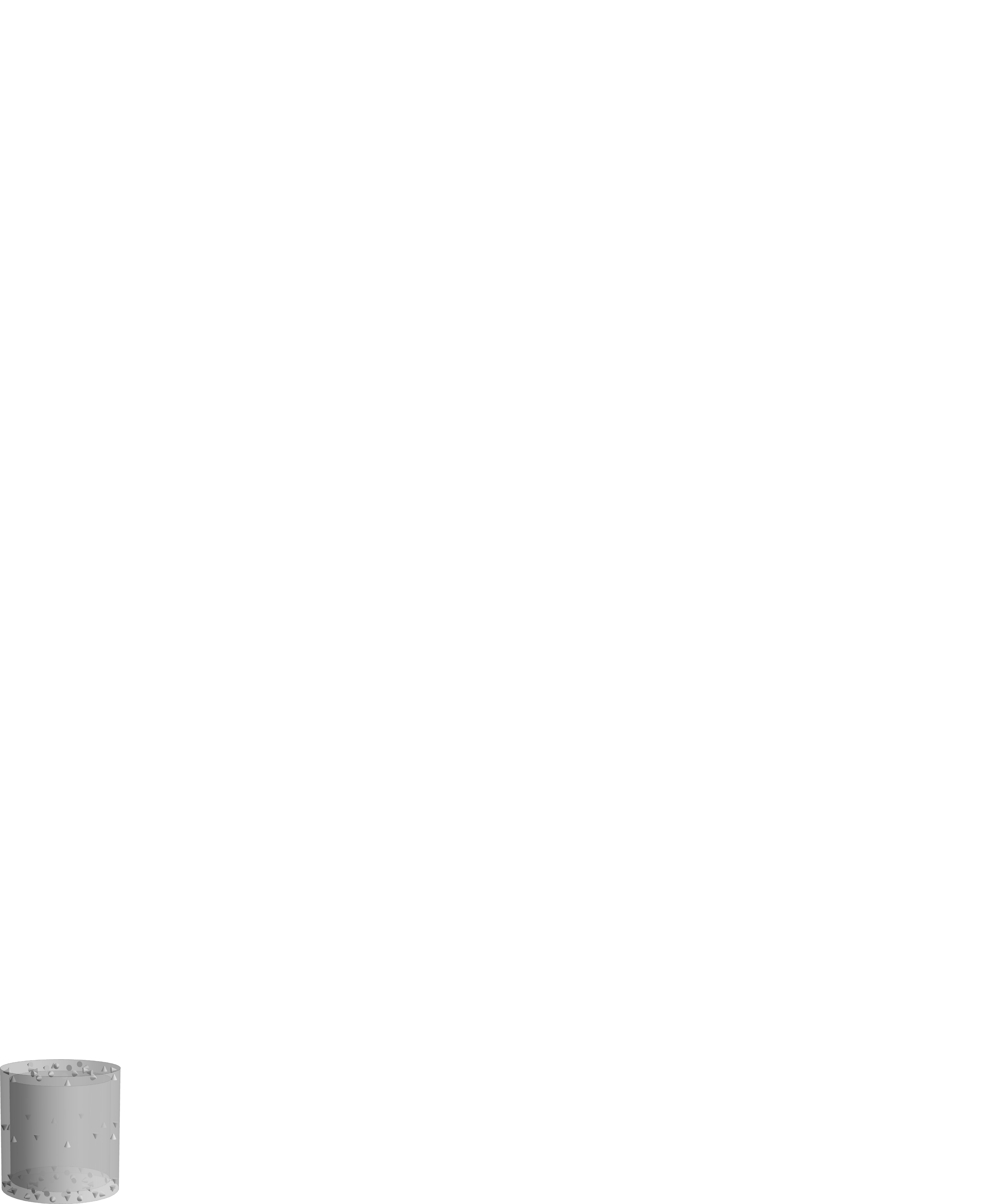}

\end{center}

\caption{\label{ivdiagram} Layout and orientation of \iv \pmts. The
cones represent the \pmts to scale, with the photocathodes being at
the large ends. The inner cylinder is the \buf vessel; detail of the
\id is not shown. The \pmts alternate directions for maximum vetoing
efficiency.}

\end{figure}

\dc is a reactor anti-neutrino experiment designed to measure the
mixing parameter $\theta_{13}$ by observing \ibd\ events, $\bar\nu_e
p \rightarrow e^+ n$. The prompt positron and delayed capture of the
neutron form the signal. The design details of the detector have been
described elsewhere \cite{Abe:2012tg}. Here, the aspects important
for muon reconstruction are given. The detector consists of four
concentric cylindrical volumes and the Outer Veto (\ov). The inner three
volumes form a single optical volume isolated from the fourth and are
collectively called the Inner Detector (\id). The four detector volumes
are:

\begin{enumerate}

\item The Neutrino Target (\targ), an innermost volume of
      gadolinium-loaded scintillator in an acrylic vessel 2.4\,m in
      height and diameter. The gadolinium is used to decrease the time
      delay and increase the observed energy of neutron capture.

\item The Gamma Catcher (\gc), surrounding the \targ, a volume of
      unloaded scintillator in an acrylic vessel of height and diameter
      3.5\,m. For muon reconstruction purposes, the \targ and \gc are
      treated as a single undifferentiated volume, as the acrylic
      separating them is only 8\,mm thick and the light output from the
      two scintillators is similar.

\item Outside the \gc is the \buf, a volume of non-scintillating oil in
      a steel vessel 5.6\,m in height and diameter in which 390 10-inch
      \pmts~\cite{pmt,pmt2,pmt3} are placed; this volume
      shields the scintillator both
      from external backgrounds and \pmt radioactivity. The \pmts
      are all aligned to point at the center of the \targ. Each \pmt
      is enclosed in a mu-metal shield resulting in a viewing angle
      of about $140^\circ$. On average, the distance from the \pmt
      photocathodes to the \gc is 0.7\,m.

\item The Inner Veto (IV), a 0.5\,m thick volume of scintillator
      outside the \buf. In this volume are 78 8-inch \pmts. These \pmts
      are arranged to maximize muon vetoing efficiency~\cite{ivpaper},
      see \fig\ref{ivdiagram}.

\end{enumerate}

Above these is the \ov, a segmented plastic scintillator detector. It
has a 13\,m$\times$7\,m lower panel of modules 1.1\,m above the \iv, and
a 7\,m$\times$3\,m upper panel 3.9\,m above the lower. Each \ov module
is made of two layers of 32 scintillator strips, either 3.2\,m or 3.6\,m
long, 50\,mm wide, staggered
by half a strip width so as to provide position information in 25\,mm
steps. In both the upper and lower \ov, two perpendicular layers of
modules are used so as to provide $x$ and $y$ coordinates. When a muon
crosses both the upper and the lower, a high-resolution track can be
reconstructed. However, since the upper \ov is much smaller than the
lower \ov, most muons that hit the \ov intersect only the lower.

Due to its overburden, 300 meters water equivalent, the \dc far detector
has a muon rate of 46\,Hz through the \iv and 13\,Hz through the \gc.
The forthcoming near detector will have a muon rate some 5 times higher.
Consequently, excellent reconstruction of muons is very helpful for
suppressing cosmogenic backgrounds. The most important of these are
\linine and \hee, which are $\beta$-n emitters. With lifetimes of
257\,ms and 172\,ms, respectively, a simple time cut after each muon
cannot be
used to remove them. They are produced at an
average distance of 500\,mm from a muon~\cite{winslow}, and therefore
can be rejected with high efficiency if the reconstruction resolution
for both the muon and the subsequent event are good enough.

A muon reconstruction can also be used to:

\begin{itemize}

\item Obtain a \dedx for muons.  This can be correlated to cosmogenic
      isotope production and used in addition to the track position 
      itself.  

\item Study cosmogenic production by stopping muons. \btw is known to be
      produced by stopping muons, but this process has not yet been
      observed for \linine or \hee.

\item Discriminate between single muon events and more complex 
      events of similar total energy. Notably, we can separate a muon
      passing through the upper corner of the \iv from an accompanying
      fast neutron interaction in the \id.

\item Image certain aspects of the detector itself \emph{in situ}. For
      instance, we used muons to verify our photogrammetric survey of
      the \ov.

\item Perform continuous timing calibration on all \pmts.

\end{itemize}

The characteristics of a muon event depend on which detector volumes
the muon intersects. Consider the case of most interest for identifying
cosmogenic isotope production, muons that pass through the \targ. The
muon, as it traverses the \id, goes through the \buf first and emits
primarily \cher light. It then crosses the \gc and \targ; while it does
so, it emits scintillator light isotropically. Finally, it leaves the
\gc and traverses the \buf again.

Muons crossing only the IV and the \buf pass through no scintillator in
the \id. The buffer oil emits \cher light and most likely a small amount
of scintillation light as well. Because of these two sources of light
and because the scintillator in the \gc absorbs the directional \cher
light and re-emits it isotropically, the overall pattern of light is
more complex than a simple \cher cone. A muon that intersects the \gc
but not the \targ may, depending on its path length in the \gc, more
closely resemble either of the above two cases.

A muon typically crosses the \iv twice. However, at the center of the
top of the \iv is the \emph{chimney} through which the \id volumes
were filled and calibration sources are inserted. About 0.5\% of muons
crossing the \id pass through enough of the chimney at the top of the
\iv to substantially reduce their \iv signal. Additionally, about 2\%
of muons stop in the \id and therefore avoid the second \iv crossing. A
muon can also cross only the \iv. All combinations are handled by the
reconstruction, with the exception of muons that stop in the \buf or
\iv.

In any of these situations, the muon may or may not cross the \ov and,
if it does, it can cross either or both \ov layers. All combinations are
handled.

Previous muon reconstruction algorithms for \dc used either the \id or
\iv data alone~\cite{Abe:2012tg,dch}. This reconstruction uses data from
all detector components simultaneously. The reconstruction algorithm
proceeds as follows:

\begin{enumerate}

\item The pulses of the \iv and \id \pmts are reconstructed as described
      in \sect \ref{sec:pulsereco}.

\item \ovexf: The event is fit ignoring \ov data using the $\chi^2$
      function described in \sect \ref{sec:func}. The fit strategy is
      given in \sect \ref{ovxfit}.

\item \ovinf: If there are \ov hits, the event is fit again with the \ov
      as a constraint using the strategy given \sect \ref{ovifit}.

\item If the event passes loose cuts for identifying a stopping muon, a
      fit under this hypothesis is done with and without use of \ov
      data. This is described in \sect \ref{stopfit}.

\item Among the fits performed, one result is chosen as the best using
      the criteria in \sect \ref{fitchoice}.

\end{enumerate}

The timing self-calibration for this reconstruction is described in
\sect \ref{selfcal}. The resolutions achieved for various scenarios are
given in \sect \ref{resolution}. Section \ref{conclusions} concludes.

\section{Pulse Reconstruction and Selection}\label{sec:pulsereco}

Each \id and \iv \pmt in \dc is read out using a 500\,MHz 8-bit
flash-ADC~\cite{fadc}. The readout window is 256\,ns. Waveforms for
all \id and
\iv \pmts are recorded for each trigger. The \id is optimized for
observation of 1--10\,MeV neutrino events rather than $\sim$1\,GeV muon
events, and so the typical waveform from a muon candidate exceeds 
the range of the \adc and does not return to baseline by the
end of the readout window. The \iv \pmts are tuned similarly to obtain
high vetoing efficiency.  To account for these facts, a special pulse
reconstruction is used for muons that is separate from that used for
other events. This reconstruction defines a start time with an error, a
rise time and total integrated charge for each pulse.

The start time of \pmts is defined as when the wave reaches halfway
between the baseline and its maximum value. If the pulse saturates the
ADC, the maximum value is simply taken to be the saturation point. The
time at which the pulse crosses the halfway point is interpolated based
on the \adc values of the two time bins that bracket the point. This
time is then corrected using the per-\pmt muon-calibrated time offsets
(see \sect \ref{selfcal}).

The rise time is defined as the time taken for the pulse to go from
10\% to 90\% of the way from the baseline to the maximum, with the same
definition of maximum as with the start time. The 10\% and 90\%
times  are also interpolated as above.

The total integrated charge for a pulse is defined as the
baseline-subtracted sum of the samples, after corrections to account for
the parts of the pulse lost due to the ADC range and the width of the
readout window. In the case of very large pulses in which the ADC is
still saturated at the end of the window, the total saturation time is
estimated using that of neighboring \pmts and the charge is assigned
accordingly.

From the rise time and charge, an error (typically between 0.6 and
2.0\,ns) is assigned to the start time of each tube. This error is drawn
from a hardcoded table which was created through an investigation of
muon fit residuals for \scindom events. A correction, as a function of
rise time and charge, is then applied to the start time. The pulses
with the smallest timing errors are those with fast rise times and
moderate charges. These are interpreted as hits free or nearly free of
\cher light. \cher light usually precedes the scintillator light and so
lengthens the rise time while making the start time less clear. Hits
with fast rise times are relatively near the muon so that they rapidly
accumulate many photons produced during the first $\sim$0.5\,ns of
scintillation. \Pmts with unusually large charges nearly always have
rapid rise times, but since their waveform is severely clipped by the
\adc it is difficult to accurately reconstruct their start time and so
have errors around 1.2\,ns.

Before the fit is run, a subset of the hits are selected for use.
The criteria for selecting these hits are different for the \id and
\iv. In the \iv, the muon may or may not be directly visible by any
given \pmt. Most \pmts will see light, but the majority of this light
is reflected. To select \pmts that see direct light, only those with
pulses that saturate the \adc and with a rise time of less than 8\,ns
are accepted. The saturation requirement selects \pmts near the muon
while the rise time requirement excludes pulses formed from several
reflections arriving with enough total light to meet the saturation
requirement.

In the \id, if the total amount of light corresponds to at least
75\,MeV deposited in scintillator, only pulses that saturate the \adc
are used. This sample of events corresponds to muons which almost
certainly intersect the \id scintillator, as opposed to \buf \cher-only
muons. Furthermore, a cut on lower energy pulses, increasing as a
function of total energy in the \id, is applied. This excludes \pmts
far from the muon which tend to worsen the fit. Overall, this selection
strategy removes the majority of hits due primarily to \cher light which
would distort the fit for events where the majority of light is from
scintillation.

When the total amount of light corresponds to less than 75\,MeV
deposited in \id scintillator, whether the muon produced more \cher
light or scintillation light depends on its trajectory. In this case,
all saturated pulses are accepted. In addition, smaller pulses are
accepted if they pass a cut which is a function of total event energy.
This cut accepts no additional pulses at 75\,MeV and pulses as small as
15\% of the saturation point at the lowest total energies. Using data,
this has been tuned for best performance as a function of energy across
the 0--75\,MeV range. At the lower total-energy end this maximizes the
statistics available for the fit while producing a clean sample of \pmts
hit by direct \cher light, excluding reflections from the vessel walls,
re-emission from the wavelength shifters in the \gc, and possible
scintillation light in the \buf. While we do not have a measurement of
how much scintillation light the \buf oil produces, the fact that this
procedure produces good fits tells us that it is subdominant to the
\cher light.

\section{Fit Function}\label{sec:func}

\begin{figure}
\begin{center}
\includegraphics[width=0.9\columnwidth]{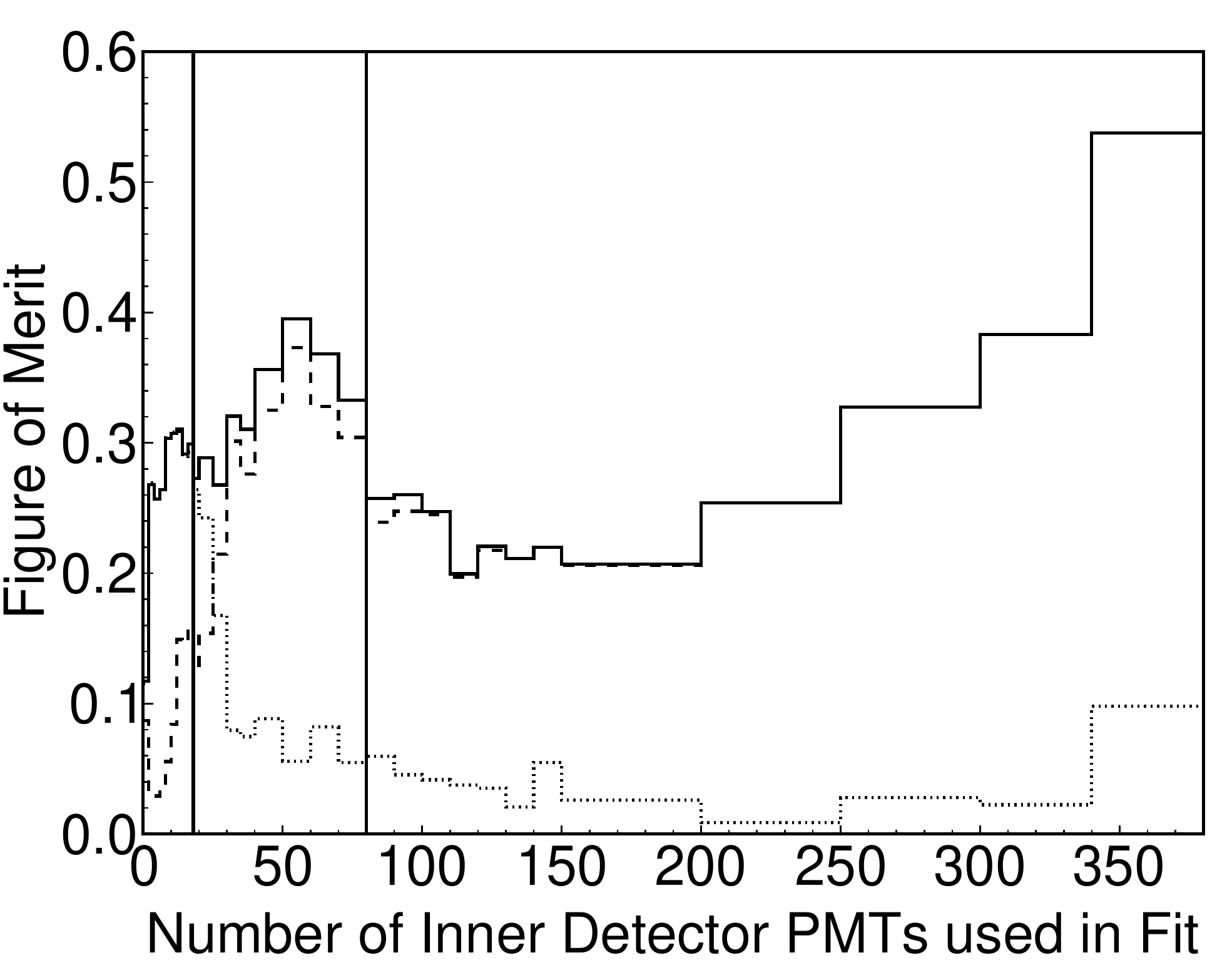}
\end{center}

\caption{\label{chervscint} Methods of choosing between the \cherdom and
\scindom models. The figure of merit is the fraction of reconstructed
tracks that intersect the \ov within 0.5\,m of the best position given
by the \ov. The dotted (dashed) line gives the figure of merit if all
events are assumed to be Cherenkov (scintillation)-dominated. The solid
line gives the case both are tried and the lower $\chi^2$ accepted. The
vertical solid lines show the cut-offs described in the text.}

\end{figure}

Both the \ovexf and \ovinf\ use the same $\chi^2$ fit function, which
\minuit \cite{minuit} minimizes. The contribution to the $\chi^2$ from
the \id is calculated either under the assumption of light mostly
from \cher or mostly from scintillation, depending on the event
characteristics. The $\chi^2$ is built from the selected \pmts' pulse
start times and their associated errors, with unselected \pmts ignored.
When \ov hits are present, they are used only as a spatial constraint
with timing information discarded. This is because the \ov uses 16\,ns
time bins, with timing significantly complicated by the long length of
the scintillator strips. In most cases, this timing would not add any
significant information.

Since one of the goals of muon reconstruction is to look for muons with
high \dedx, we do \emph{not} use the total reconstructed energy in the
fit function for through-going muons. There are a few exceptions to this
which will be described later.

\subsection{Choice of \id Model}\label{fitass}

The first step of the reconstruction is to decide whether to assume that
the \id is dominated by scintillation light, because the muon traversed
a significant length of the \gc, or instead dominated by \cher light.
Different fit $\chi^2$ functions are used for the \id in these two
cases. The \iv is handled the same way in either case.

If the number of selected \id \pmts is fewer than 18 ($\lesssim$30\,MeV
in scintillator-equivalent light), the code assumes a \cherdom event. If
there are more than 80 ($\gtrsim$180\,MeV), instead the event is assumed
to be \scindom. Between these, it tries both hypotheses and chooses the
one that produces the better $\chi^2$. As shown in \fig\ref{chervscint},
the $\chi^2$ effectively chooses the better solution. These cut-off
values were chosen to cover the cross-over region without wasting time
trying hypotheses unlikely to be chosen.

\subsection{\id Model for Scintillator-dominated
Events}\label{sec:scintdom}

\begin{figure}

\begin{center}

\includegraphics[width=0.72\columnwidth]{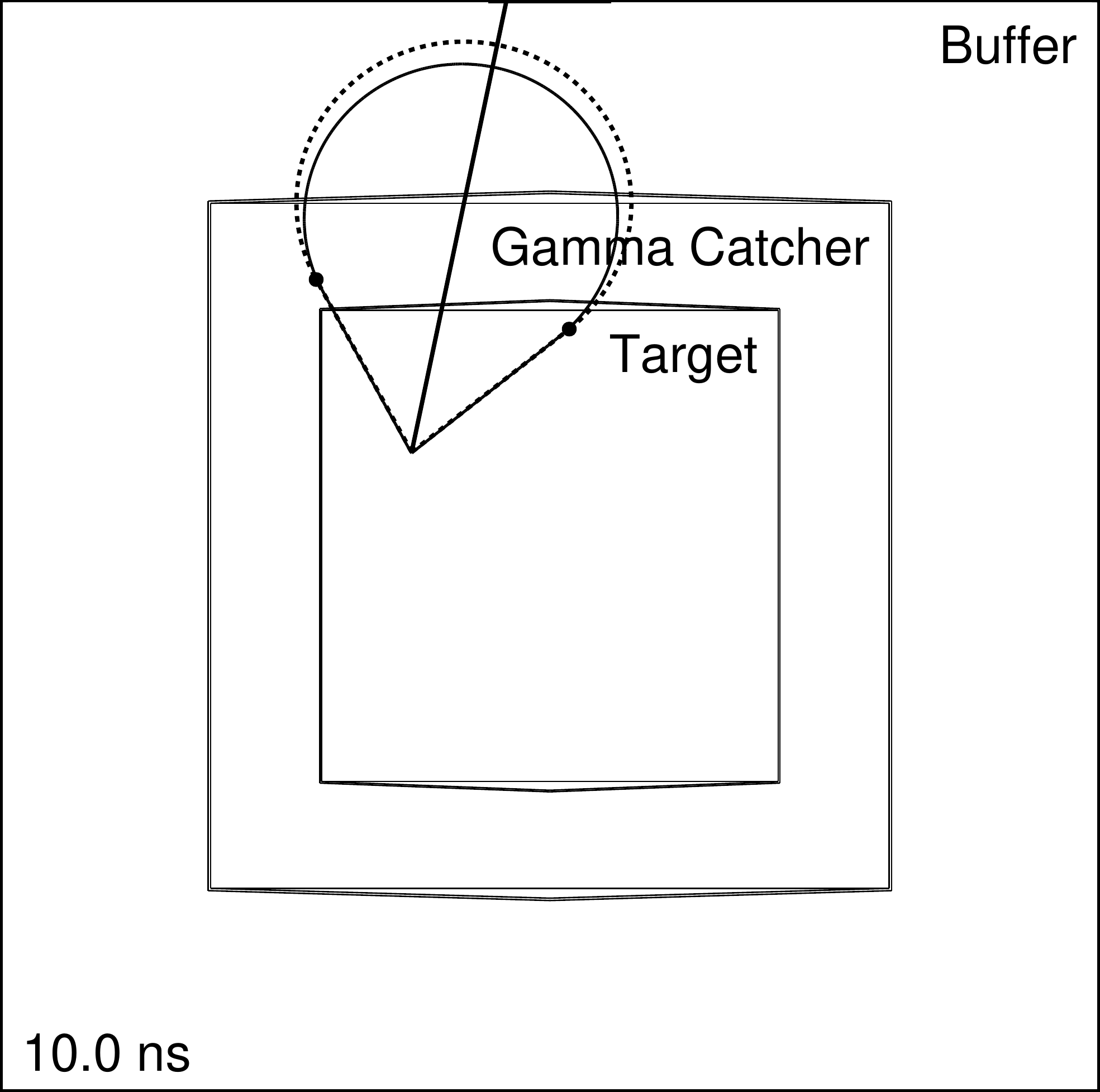}

\includegraphics[width=0.72\columnwidth]{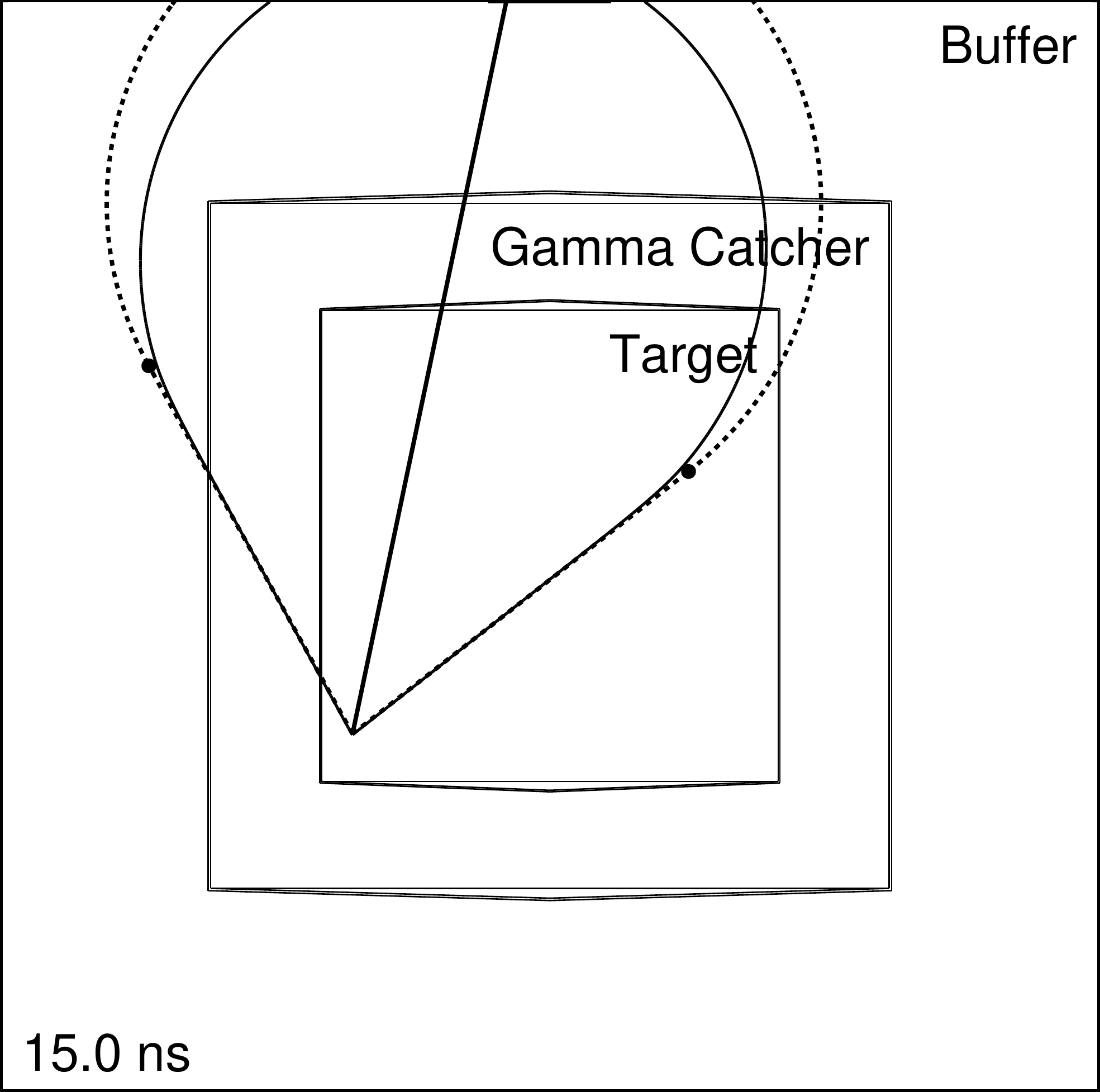}

\includegraphics[width=0.72\columnwidth]{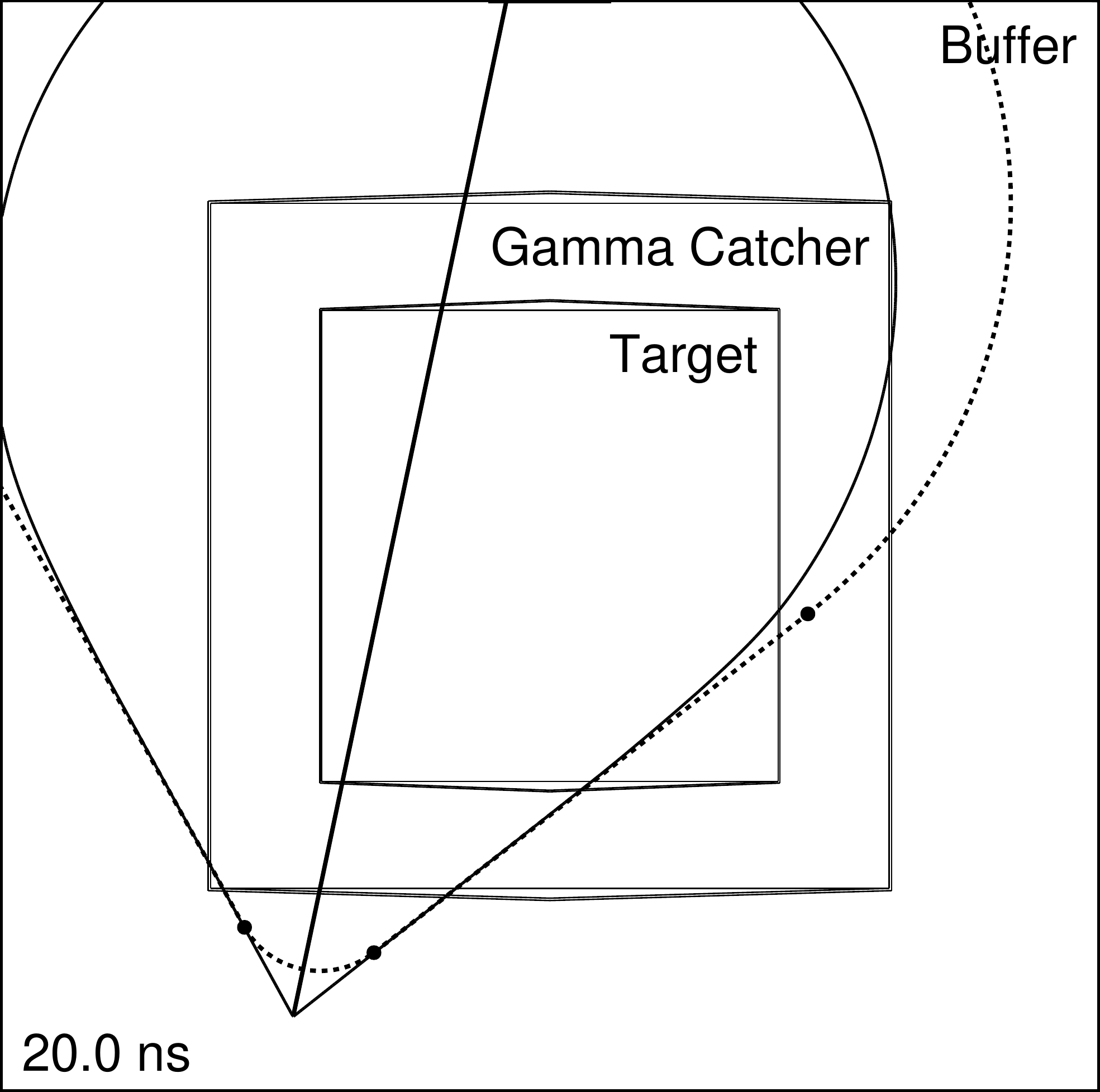}

\end{center}

\caption{ \label{lightcartoon} The wavefront of scintillator light
emitted by a muon as it traverses the \id. The dashed line denotes the
earliest scintillation photons. The transitions between the conical and
spherical wavefronts are marked with black dots. The solid line gives
the effective wavefront as used by the reconstruction. As described in
the text, this lags the first scintillation photons in the region above
the scintillator, and leads it below where \cher light is significant.}

\end{figure}

The \targ and \gc form a cylinder of scintillator. Extending from the
top of the \gc is the chimney, two concentric clear acrylic tubes, with
outer radius 188\,mm, containing \gc and \targ scintillator. While
light from the chimney scintillator can contribute to the observed
event, it is rare for a muon to pass though a significant amount of it.
It is therefore ignored for purposes of muon reconstruction. The \id
scintillator is modeled as a simple cylinder with the dimensions of the
\gc.

Given a particular muon trajectory, we calculate when each \pmt's
pulse start time should be and compare to the observed time to form
the $\chi^2$. As the muon traverses the scintillator it emits light
isotropically. Due to the muon's motion, this light forms a cone similar
to that of \cher radiation. In addition to the cone, a sphere of light
is produced behind the point where the muon enters the scintillator. The
same occurs at the exit point. This is shown in \fig\ref{lightcartoon}.
There are, then, three regimes to consider:

\begin{enumerate}

\item[(1)] The \pmt sees light first from the point at which the muon
enters scintillator. As can be seen in \fig\ref{lightcartoon}, a large
fraction of the \pmt's typically fall in this category.

\item[(2)] The first light comes from some point along the muon's track
through the scintillator, i.e.\ the scintillation cone intersects these
\pmts.

\item[(3)] The \pmt's first light comes from the point at which the muon
exits the scintillator.

\end{enumerate}

A clean separation of \pmt's into these three categories is an
idealization which turns out to be too far from the truth to be used
without modification. It ignores \cher light and assumes arbitrarily
bright scintillation. Two changes are therefore made:

First, while \cher light is generally ignored for this category of
events, the \pmts near the muon's exit point are very close to the muon
and see much more of this light than others. Their observed timing is
often due to \cher light alone. As a reasonable approximation, we simply
continue to model the cone of light after the muon has left the \gc,
i.e.\ case (3) is collapsed into case (2). Since this affects only on
order of 10 \pmts in most cases, no attempt is made to adjust the cone
parameters to reflect the different characteristics of \cher light. A
similar detector with a higher density of \pmts would benefit from an
effort to resolve the \cher and scintillation light separately in this
region.

The second change is more involved and is necessary due to the
sharp boundary between cases (1) and (2) above. This sharpness is
physically unrealistic, since it implies that an infinitesimal volume
of scintillator at the entry point produces enough light to be seen by
every \pmt that lies above the region intersected by the scintillation
cone. In reality, each tube must have received light from a significant
length of the muon track before it crosses the start-time threshold. To
model this, we compute an \emph{effective} position in the scintillator
for each tube that represents where its first light was produced. As
compared to the point in the scintillator that represents the earliest
possible light seen by a given tube, this effective position is shifted
away from the edge of the scintillator if the ideal position is at or
close to the edge. The farther the ideal position is to the edge, the
smaller the shift. If the ideal position is sufficiently deep in the
scintillator, no shift is applied.

The functional form used to calculate these shifts is not derived from
first principles, but simply designed to make a smooth transition
between case (1) in which a shift is needed, and positions deep in the
scintillator where we do not modify the position. It was chosen to be
both reasonably fast to compute and to result in start time predictions
that are generally differentiable with respect to all track parameters
so that \minuit does not see sudden changes in the $\chi^2$ function.

Let $\smalllength$ be the shift applied in case (1) in which the
ideal first-light point is the edge of the scintillator. This is the
largest shift that we impose. This length is proportional to the square
of the distance between the scintillator edge and the \pmt. We will
apply a shift between $\smalllength$ and zero if the ideal position is
between zero and $L$ from the edge of the scintillator, where $L = \pi
\smalllength / (\pi - 2)$. The new position is \[ x' = \smalllength + (L
- \smalllength)\left(1-\cos\frac{\pi x}{2L}\right), \] where $x$ is the
distance from the edge of the scintillator to the ideal position. If the
ideal position is farther from the scintillator entry point than $L$,
the position is unmodified.

If $L$ is greater than half of the length of the track through
scintillator, $s$, the procedure is modified to prevent the effective
position from being shifted past the halfway point: \[ x' = \smalllength
+ \left(\frac{s}{2} - \smalllength\right) \left(1-\cos\frac{\pi
x}{s}\right), \] unless $\smalllength$ is also greater than half of $s$,
in which case, we use simply \[ x' = \frac{s}{2}. \]

Because muons of different energies produce different amounts of light
per unit scintillator length, there is no single correct value of the
proportionality constant, the \emph{transition length parameter}, that
determines $\smalllength$. Higher energy muons have a higher \dedx and
therefore a shorter transition length. It can also be substantially
shorter if a muon showers near the top of the detector or the muon-like
event is actually two closely spaced muons from the same air shower.
Therefore, this parameter is allowed to float in the fit. In typical
situations, $\smalllength \approx 300$\,mm, which delays the modeled
\pmt start time by $\sim$1\,ns.

\begin{figure}

\begin{center}
\includegraphics{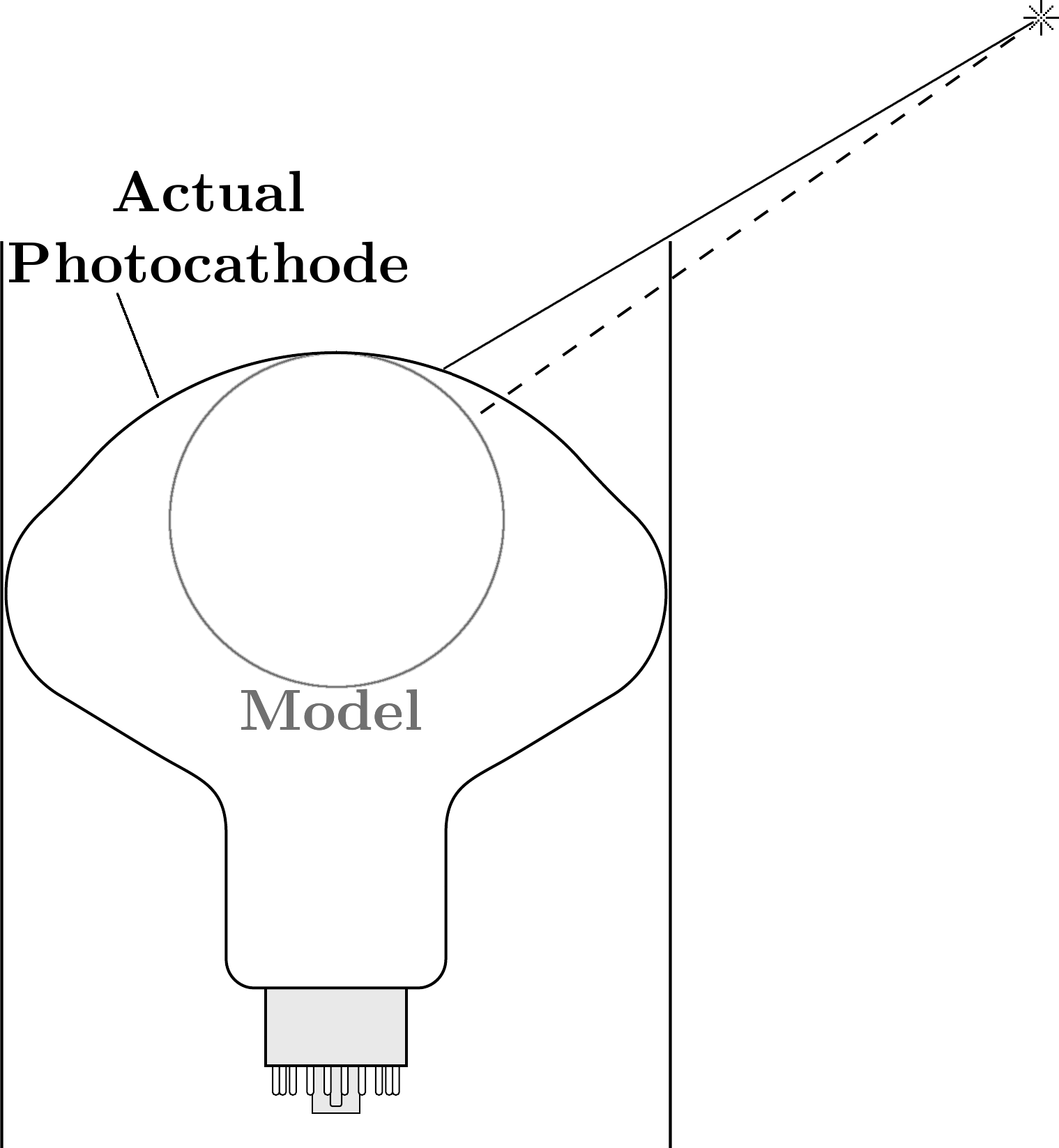}
\end{center}

\caption{\label{idpmt} \id \pmt with mu-metal shield.  The gray circle
shows the model of the photocathode used by the reconstruction to 
produce approximately the correct timing without the need to explicitly
model the shield.  For an example source of light shown in the upper
right, the solid line shows the true path of light to the photocathode.
The dashed line shows the modeled path, which is of nearly the same
length.}

\end{figure}

Once the effective position of the first light emission seen by the
\pmt is calculated, the expected \pmt start time is calculated. Because
we can achieve precision similar to the size of the \pmts, modeling
them as point objects is insufficient. We instead make an approximation
that keeps the computational cost low while accounting for both the
size of the photocathode and the viewing angle available given the
mu-metal shields. As shown in \fig\ref{idpmt}, the photocathodes are
modeled as spheres of half the actual radius of the photocathodes,
and the shields are not explicitly modeled. The centers of the model
photocathodes are aligned with the centers of the real photocathodes. It
is implicitly calculated where on this model of the photocathode light
will strike first by calculating the time for light to reach the center
of the sphere and then subtracting off the radius. This treatment has
nearly the same computational cost as modeling the \pmt as a point while
consistently handling a variable speed of light, as discussed below. It
gives substantially better resolution than approximating the PMT as a
point at the center of the photocathode, or using a sphere with the full
radius of the photocathode.

As with the transition length parameter, there is no single correct
speed of light. The speed is a function of wavelength, and so is
affected by the initial spectrum produced in each detector liquid,
the reemission of this light by wavelength shifters in either the
same volume or another, the wavelength-dependent attenuation, and
the wavelength-dependent \pmt sensitivity. Therefore, it is allowed
to float in the fit, although it is constrained to stay within a
reasonable range. This produces a single effective speed of light for
the particular track being reconstructed. It should be noted that the
effect of this speed parameter is primarily to adjust the opening
angle of the scintillation cone. Because the apparent opening angle is
distorted by the variation in light intensity seen by the various \pmts,
even after corrections described in section \ref{sec:pulsereco}, this is
not an effective way to measure the true speed of light.

In all, there are seven free parameters in the \scindom fit: four
spatial parameters that define the track itself, the muon entry time,
the transition length and the speed of light.

\subsection{\id Treatment for \cherdom Events}\label{sec:cherdom}

The \cher fit is substantially simpler. While the complete pattern of
\cher light is complicated due to absorption and re-emission in the \gc,
by having selected only the \pmts with the highest charge using the
procedure in \sect \ref{sec:pulsereco}, usually there is only a small
line or disk of \pmts on one wall of the \id that have seen direct,
intense \cher light. The threshold for accepting pulses has been tuned
using the data to optimize the resolution by using as many pulses as
possible without accepting ones due to indirect light.

The expected \pmt start time for each \pmt is calculated by finding
the position along the track from which \cher light directed at the \pmt
was emitted, initially ignoring the boundaries of the detector. The
initial $\chi^2$ is based on this time. We then check, for each \pmt,
that the light is coming from in front of the tube. Since, unlike the
scintillator case, the light is directional, if this is violated, the
\pmt should not see any light at all. If the light is coming from the
wrong direction, a $\chi^2$ penalty term is applied that is zero if the
light is coming in orthogonal to the \pmt axis and rises as the square
of the cosine of the angle as the direction becomes more backwards. In
this way, the track is steered into the correct general location so that
the main timing component of the $\chi^2$ can be usefully minimized.

To constrain the track to the correct region of the detector, the
\cherdom fit also has two additional overall penalty terms, one that
prevents the track from leaving the \buf entirely, assuming that at
least one \id \pmt was selected for the fit, and another that prevents
it from crossing an unrealistic length of \gc scintillator. Since a
significant amount of scintillator can be crossed while leaving the
event with primarily \cher-like characteristics, this second penalty
does not begin until 500\,mm of the \gc is crossed.

The \cherdom fit does not allow the speed of light to float. Because of
the small number of \pmts used, and the close proximity of these \pmts
to the muon track, the fit would not sufficiently constrain the speed.
It therefore has only five free parameters.

\subsection{\iv Model}\label{sec:ivpart}

Under either fit hypothesis, \iv timing information is used in the same
way. Since the \iv is relatively thin, an approximation is used in which
the scintillation light is treated as coming from a single point for the
muon's entry position and another for its exit. These points are found
by computing the intersection of the track with a cylinder that lies
halfway between the \buf vessel and the \iv vessel. If the track does
not intersect this cylinder because it only clips a corner or grazes
the edge, a single point is used instead, centered on the intersection
with the \iv. For simplicity, the smaller \iv \pmts are modeled as
point objects with a \pmt's position taken to be the center of its
photocathode.

The first light observed by a \pmt can be either from the muon's
first or second crossing of the \iv. The code considers each
possibility and chooses the one that has a better $\chi^2$ including
these contributions:

\begin{itemize}

\item The $\chi^2$ from timing, taking the expected arrival time of
      light to be the simple distance from the light production point to
      the \pmt.

\item A penalty term that adds to the $\chi^2$ if the line connecting
      the light production point to the \pmt goes through the \buf
      vessel. This penalty is a function of the length of \buf
      intersected and rises smoothly from zero for zero length.

\item A penalty term that checks whether the light comes from behind
      the \pmt. Similarly, this penalty rises smoothly from zero as the
      angle between the \pmt axis and the light direction increases from
      $90^\circ$.

\end{itemize}

Usually the fit does not explicitly constrain the track to be near the
selected \pmts. This would, in general, be detrimental because the \pmts
are not uniformly distributed and because the way that \pmts are chosen
does not necessarily select the nearest ones. However, when the total
number of \pmts in the fit falls below the number of free parameters in
the fit, proximity to \iv \pmts is added as a fourth penalty term. In
this case an additional $\chi^2$ term is also used that constrains the
track to have a length in the \iv scintillator consistent with the total
\iv charge, assuming a minimum ionizing muon.

Additional penalties are applied if the track crosses the \id volume in
a way severely inconsistent with the signal in the \id, for instance if
no \pmts were selected for use in the \id but the track passes through a
region easily visible to them, or if the track only crosses a very small
amount of the \iv. These penalties are used primarily to guide \minuit
into a reasonable parameter region and ideally are all small or zero for
the final result.

\section{Fit Strategy}

\subsection{\ovexf}\label{ovxfit}

The tracks are parameterized by their position as they enter and
exit the \iv and their time when they enter the \iv. The positions
are represented by their polar and azimuthal angles, $\theta$
and $\phi$, with respect to the center of the detector. For the
scintillator-dominated fit, the transition length parameter from \sect
\ref{sec:scintdom} and the speed of light are also free parameters.

\minuit is ultimately used to minimize the fit function, but cannot
be relied upon to find the correct minimum from an arbitrary set of
initial parameter values. Instead, to get a rough idea of where a track
should be, the code first tests a table of uniformly distributed track
guesses. For each guess, \minuit is used to minimize the $\chi^2$
allowing only the muon entry time to vary. 194 guesses are used for
\scindom events and 452 for \cherdom events, where these numbers result
from making uniform steps in $\theta$ and $\phi$ and selecting the
results that intersect the correct detector volumes. If the density of
these guesses is too low, the final results can erroneously cluster
near the guesses themselves. With the density chosen, this effect
is only significant when the number of selected \pmts is below 10.
Although all of the guesses are \downgo, \minuit will, in the next step,
not use this as a constraint. A small number of tracks are therefore
reconstructed as \upgo. The user can alternatively ask for an equal
number of \upgo guesses to be used. However, the rate of \upgo muons in
\dc is negligible so this is not done by default.

The result with the lowest $\chi^2$ is used as input into a full fit.
Even though at this point we expect to be near a local minimum in the
fit function, and some attempt has been made to make the function
smooth, it is nevertheless not smooth everywhere. We have found that
\minuit often fails to find the minimum via a single call to \migrad.
Therefore the following progression of \migrad calls are used. First,
all entry and exit angles are limited to $[-4\pi, 4\pi]$, which prevents
pathologies when, for instance, $\theta$ is near zero while still giving
\migrad quite a bit of freedom to choose its path towards the minimum.
In contrast, if we were to limit $\theta$ to $[0, \pi]$ and $\phi$ to
$[-\pi, \pi]$, it would be difficult to find minima near one of the
limits. If \migrad does not immediately report success, it is called
with successively higher values of its ``tolerance'' parameter until it
does. This sequence helps negotiate kinks in the $\chi^2$ function due
to physical boundaries such as moving the track from the lid of the \gc
to the side wall. Assuming this is achieved, the angle limits are lifted
and \migrad is called once more as recommended by Ref.~\cite{minuit}. If
the best $\chi^2$ happens to be a in a well-behaved region, errors are
extracted using {\sc hesse}.

Finally, if the $\chi^2$ is more than three times the number of
degrees of freedom or the \dedx in the \id of the resulting track is
significantly lower than minimum ionizing, the entire procedure is
repeated with a higher density of initial guesses in an attempt to find
a better solution.

\subsubsection{\cher Removal}

When doing a fit under the \scindom hypothesis, there may nevertheless
be \pmts that cross the threshold before scintillation light hits them
due to being near the muon entry or exit point and seeing a large amount
of \cher light. Near the exit point this is a mild effect, as the
scintillation light and \cher light arrive only a few nanoseconds apart,
and the treatment of case (3) in \sect \ref{sec:scintdom} mostly takes
care of the problem. However, \pmts near the entry point can see \cher
light much earlier than scintillation light since the \cher light comes
directly from the muon as soon as it enters the \buf.

After doing the full fit procedure, the code checks for such \pmts. If
found, they are removed and the fit is repeated. This is done up to two
times, with up to three \pmts being removed from the fit each time.
\pmts are deemed to be \cher contamination if they are at least 10\,ns
early compared to the fit expectation, or alternatively if they are at
least 7\,ns early, within 2\,m
of the track, and closer to the entrance in the \buf than the exit.

This algorithm also eliminates other classes of early hits such as \pmt
pre-pulses and accidental coincidences with non-muon processes.

In the case of an intermediate amount of light (see \sect \ref{fitass})
in which the choice between \scindom and \cherdom is based on the fit
results, the \cher removal is only performed after this decision has
been made. When there is \ov data, \cher removal is only done during the
\ovexf and the removed \pmts remain removed during the \ovinf. This is
for performance reasons only and not a necessary feature.

\subsection{\ovinf}\label{ovifit}

If both the upper and lower panels of the \ov provide \xyo, \emph{\ov
tracks} are formed. Typically more than one \ovt is formed from a
single muon due to crosstalk, bremsstrahlung, showering, etc. Since
these tracks are very high resolution, we take it as given that one
of them is correct and our only task is to choose between them. \ovts
can be rejected, however, if they miss the rest of the detector, or
are severely inconsistent with the \id or \iv signals. Each acceptable
\ovt is tested in turn by doing a fit that varies only the non-spatial
track parameters (entry time, speed of light and transition length,
as applicable). Using a combination of the resulting $\chi^2$ and the
quality of the \ovt, it selects one.

Usually, no \ovts are present (or acceptable), but one of the two \ov
panels provides several possible sets of \xyo. The 6 of these with the
largest energy deposition are considered for inclusion in the fit.
Each is tested by repeating the fit with the track constrained to pass
through the scintillator strips involved.

In some cases, the muon is only registered in one \ov module and so
the \ov position is known very poorly. In this case, the fit is done
with the loose constraint that the track must pass somewhere through
the \ov. While most low-quality \ov triggers of this sort result from
radioactivity rather than muons, the fraction of these fits resulting
from accidental coincidences is only 0.2\%.

As with the \ovexf, a variety of initial conditions are tried, including
both the result of the \ovexf, adjusted as appropriate, and a fixed
table of uniformly distributed guesses. For use by \minuit, the spatial
component of the track is parameterized by the $x$ and $y$ position
within the \ov and the $\theta$ and $\phi$ of the track. In the $\chi^2$
function, this is translated into the representation described in
section \ref{ovxfit} so that no code is duplicated.

As above, the code is usually configured to assume that all muons are
\downgo, but can also attempt \upgo if requested. In this case, the
entire procedure is done each way and the overall result with the lower
$\chi^2$ is returned.

\subsection{Stopping Muons}\label{stopfit}

If the muon deposited an unusually low amount of energy in the \iv, it
is a stopping muon candidate. All muon-like events with less than 70\%
of the mean muon \iv energy are reconstructed under both \thrugo and
stopping hypotheses. Because it is not possible to know the expected
amount of light produced in the \iv without first knowing the muon
trajectory, this is a loose cut that certainly covers all muons that
stop in the \id while still excluding most \thrugo muons to save
processing time.

The stopping muon fit function that \minuit minimizes is the same as the
\thrugo \scindom fit, except that:

\begin{itemize}

\item Light is only expected from the entry point into the \iv rather
      than at two points.

\item An additional free parameter is added to the fit: the fraction
      of the \gc and \targ crossed along the track trajectory. No light
      is produced past this point.

\item The method described in \sect \ref{sec:scintdom} to smooth out the
      scintillator entry points is applied to both the entry point
      and the stopping point, since there is no \cher light from the
      \buf after the exit point if the muon does not exit. The muon
      track becomes visible to the \pmts below the stopping point much
      more rapidly than to those above the entry point due to the
      scintillation light piling up behind the muon. Because of this, a
      shorter transition length is used at the stopping point.

\item The track length in the \id is strongly constrained to match the
      observed \id energy using a $\chi^2$ penalty term. Unlike \thrugo
      muons, stopping muons do not have enough energy to shower and
      should have a one-to-one correspondence between track length
      and visible energy.

\end{itemize}

The muon is still assumed to travel at the speed of light up to the
stopping point. This is a good approximation: only in the last
250\,mm does it drop below 250\,mm/ns.

The overall fitting strategy is the same as for the \thrugo \scindom
fit, except that the speed of light and the transition length parameter
are not allowed to float since allowing them to float was found to
worsen the fit results. The same density of initial track guesses is
used. For the initial fits in which only the muon entry time is allowed
to vary, the stopping position parameter is fixed at the value that
produces the right scintillator path length for the observed energy.
This fit is done both with and without the \ov if \ov data is present.

\subsection{Performance}

On a 2.4\,GHz Intel Xeon CPU, the mean time used per muon is 0.12\,s.
At the \dc far detector rate, this means about 6~CPU-hours are needed to
reconstruct 1~hour of data. About half of the runtime is used doing
one-parameter time-only fits to determine the correct starting point for
each type of fit, as described in \sect \ref{ovxfit}. The other half is
used doing the full fits from the favored starting points.

\section{Choice of Fit} \label{fitchoice}

As described, up to four fits are done --- all combinations of
\ov-ex\-clu\-sive/in\-clu\-sive and \thrugo/stop\-ping. One is selected
as the best answer. For the \thrugo fits, if there is OV data, the
\ovinf is usually judged to be better than the \ovexf on the basis
that the \ov position information is relatively unambiguous. However,
since it is possible for the \ov signal to be due to an accidental
coincidence, or for the \ov hits to be from secondaries rather than the
muon itself, from other muons originating in the same air shower, or for
there to be a pathology in the \ov-inclusive reconstruction, the \ovexf
is selected if its $\chi^2$ is at least 1000 units lower.

\begin{figure}

\begin{center}
\includegraphics[width=0.9\columnwidth]{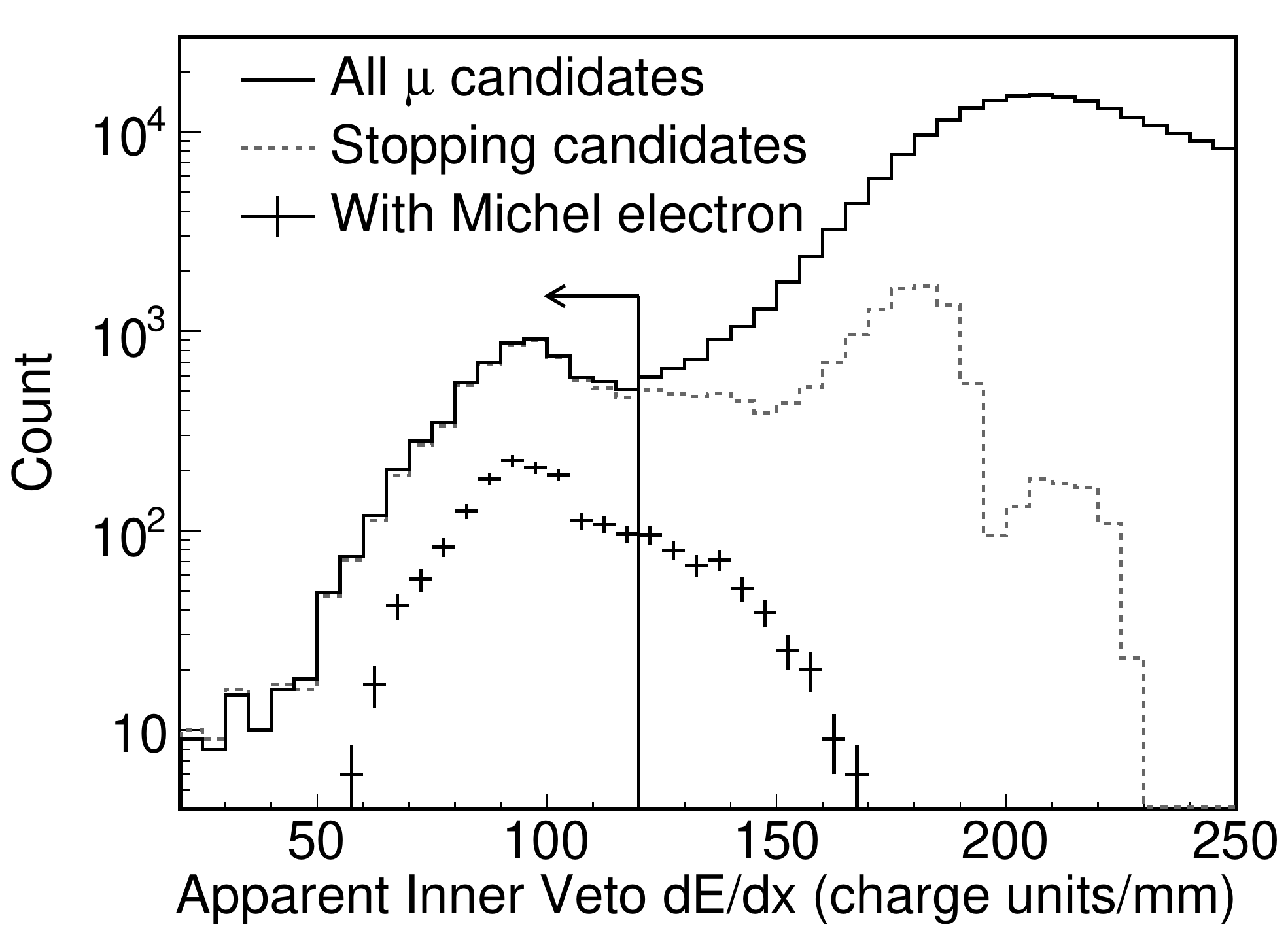}
\end{center}

\caption{\label{fig:stoppingcut} \dedx in the \iv for muons fit under
the \thrugo hypothesis. The solid line shows all muons. \Thrugo minimum
ionizing muons make up the broad peak around 210. The dashed line
shows muons fit under the stopping muon hypothesis. The stopping muon
hypothesis is favored if \dedx $< 120$. Finally, muons followed by
Michel electrons are shown. The efficiency for observing these is low.
In particular, electrons produced in the \buf are lost.}

\end{figure}

If a stopping muon fit was done, it is chosen as the best answer
if the \iv energy is consistent with a single \iv crossing, given
the reconstructed length of \iv scintillator traversed (see
\fig\ref{fig:stoppingcut}). If \ov information is present and the
stopping fit is chosen, the \ovinf is always chosen. There is
no option to reject this fit in favor of the \ovexf because
the low statistics of the stopping muon sample make it difficult to
determine when, if ever, this would be beneficial.

While one of the fits is chosen as the best, all four fit results
are also saved separately. This is useful since some analyzers may
wish, for instance, to use the \ovexf more often to exclude accidental
coincidences, or, in the case of \thrugo versus stopping, one may wish
to apply a different cut to obtain either a larger sample or a more pure
one. It also allows the reconstruction to be tuned by comparing the
agreement of the {\ovexf}s to the \ov data.

\section{Self-calibration}\label{selfcal}

\dc has two light injection systems used for timing calibration,
a multiwavelength LED-fiber system installed on the buffer
walls~\cite{Abe:2012tg}, and a 470 nm laser diffuser which is deployed
along the vertical axis.
The latter is more precise, producing time constants with an
uncertainty of 0.15\,ns; the two systems are found to give consistent
results within 0.5\,ns. We find that with the calibrations derived
from these systems applied, the per-\pmt muon fit residuals are
typically offset by up to 1\,ns. The muon statistics available in each
hour-long run are sufficient to measure these offsets with a precision
of 0.02\,ns. These offsets are believed to be a consequence of the
reconstruction's imperfect model of muon light distribution coupled
with the fact that different \pmts tend to preferentially sample
different parts of the distribution.

To correct for this, the residuals are used to produce a new set of
timing calibrations using an iterative procedure. Because the \dc far
detector is under a hill, the muon flux is not uniform as a function
of the azimuthal angle. Calibration events are weighted so that all
azimuthal angles are equally represented. If we do not perform this
weighting, the fit resolution suffers and the fit tracks are biased
towards the direction of the average muon.

The self-calibration is done for each 1-hour run. It improves the
resolution by 16\% while also providing the best measure of \pmt timing
stability for \dc, since no dedicated calibration runs are needed.

\section{Resolution}\label{resolution}

The resolution of the fit is tested using data-driven methods, primarily
in events with \ovts. The resolution of the \ovexf can be tested by
comparing these fits with \ovts. In the same way, the resolution of fits
done with \xyo in a single \ov panel can be tested. While the acceptance
of \ovts is small due to the limited size of the upper \ov, the symmetry
of the \id suggests that resolutions measured using this method are
generally valid. One can also compare \ovexf fits to hits in the lower
\ov panel alone when \ovts are not available. This provides limited
information, but a much larger acceptance. Results from this method
agree with that of \ovt comparison.

\begin{figure}\begin{center}

\includegraphics[width=0.9\columnwidth]{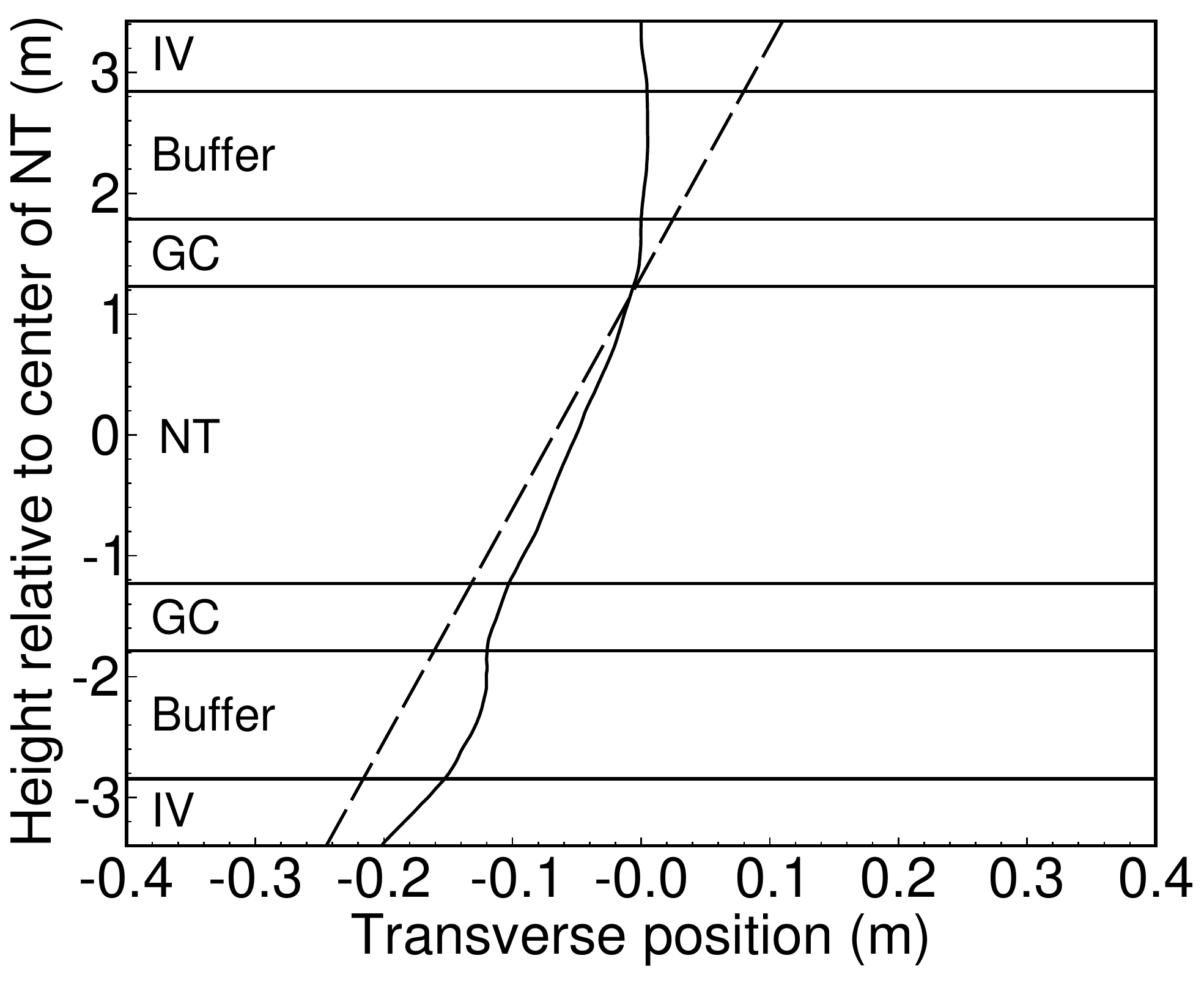}

\end{center}

\caption{\label{fig:scattering} Reconstruction of a low energy muon.
The solid line is a simulated muon track.  The muon initially has
a kinetic energy of 1.5\,GeV and is going straight down.  It scatters
through 0.2\,m as it traverses the detector (note exaggerated horizontal
scale). The dashed line shows the reconstructed track. }

\end{figure}

If the resolution were limited by \pmt timing alone, we would expect,
in the ideal case of a minimum ionizing \thrugo muon intersecting a
large amount of \id scintillator (8\% of the reconstructed muons), to
obtain a resolution of about 20\,mm in each transverse coordinate at the
detector center. In other words, this is the error that \minuit reports.
At low muon energies, this is clearly unobtainable since the fit assumes
that muons travel in straight lines, while real muons undergo multiple
scattering (see \fig\ref{fig:scattering}). The median muon energy at
the \dc far detector is about 30\,GeV; at this energy the RMS deviation
due to multiple scattering as a muon crosses the \id is 40\,mm in each
transverse coordinate.%
\begin{figure}

\begin{center}

\includegraphics[width=\columnwidth]{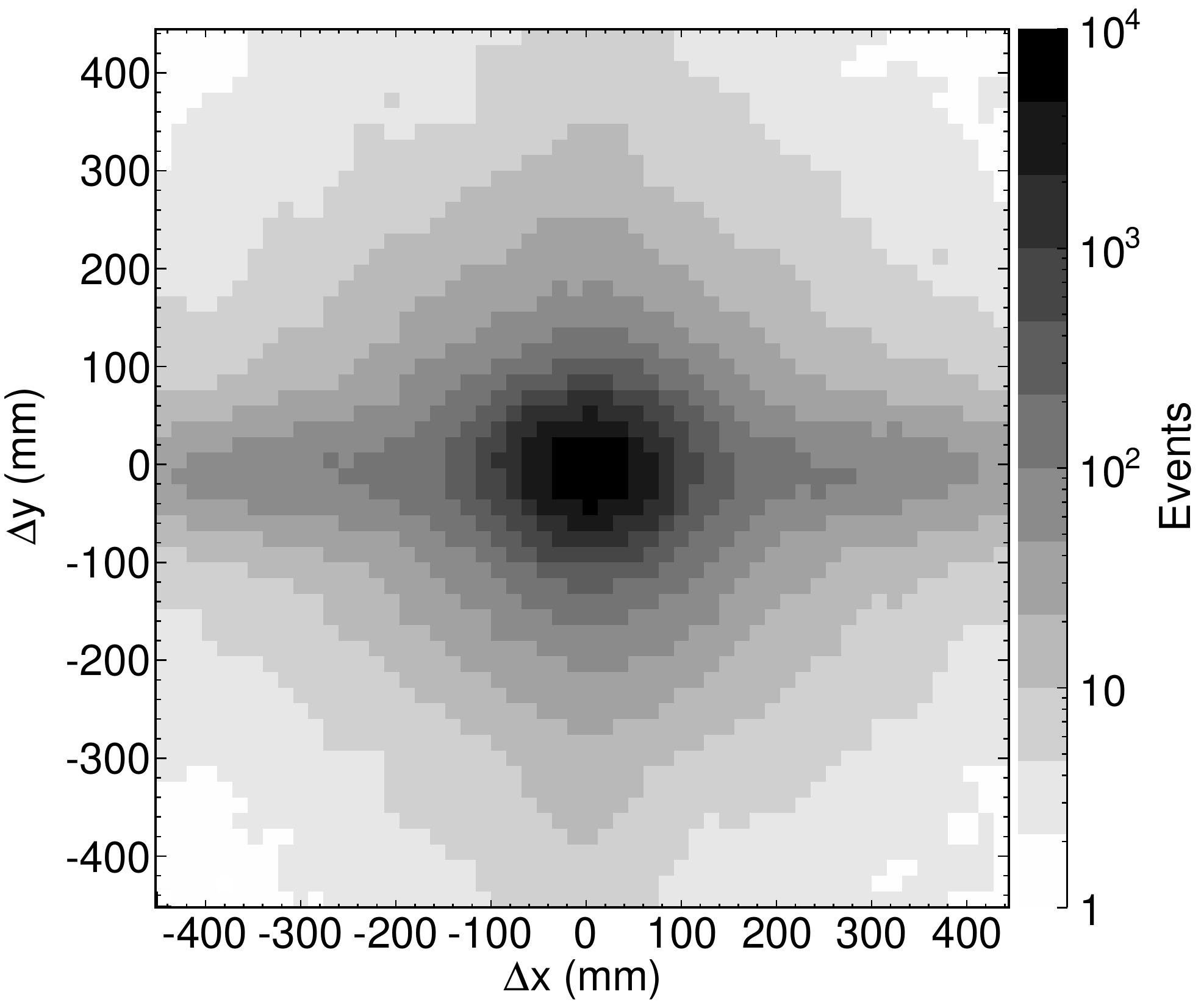}

\end{center}

\caption{\label{fig:resolution} Resolution as measured by \ovts. Shown
is the relative position in the $z=0$ plane between the \ovt and the
result of this reconstruction without use of \ov data. Only events in
which all \id \pmts are used are plotted here. This \ov's own resolution
is a significant effect. The ridge along $\Delta y = 0$ arises from \ovt
misreconstructions and the asymmetry of the \ov.}

\end{figure}\begin{figure}

\begin{center}

\includegraphics[width=0.95\columnwidth]{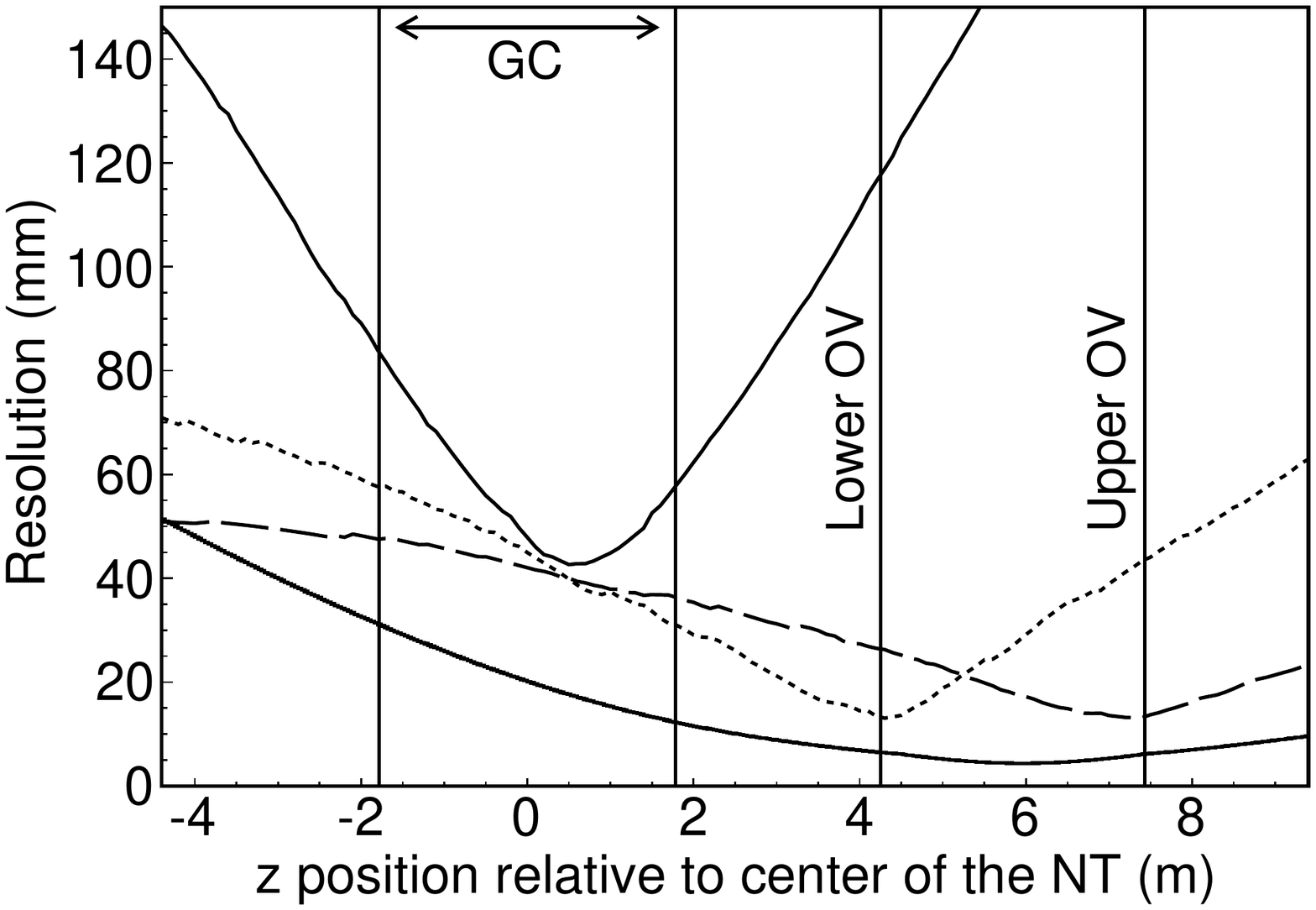}

\end{center}

\caption{\label{fig:resbyz} Resolution along reconstructed tracks
as a function of vertical
position for events with \ov tracks and all \id \pmts used. The upper
solid line gives the resolution if the \ov is ignored. The dashed
(dotted) line gives the resolution if only the upper (lower) \ov is
used. The lower solid line is the resolution of the \ov track. The
vertical extent of the \gc and positions of the \ov panels are shown.}

\end{figure}
By comparison with \ovts, we find that the resolution is about 40\,mm at
the center of the detector in this case (see \fig\ref{fig:resolution}).
The resolution is larger at the top and bottom of the detector by about
a factor of 2, as shown in \fig\ref{fig:resbyz}. This figure compares
the resolution if the \ov is not used to that obtained with either the
upper or lower \ov panel alone, and to the resolution of \ovt. \ov
hits are present in about half of events across all categories of
muons, and \ovts in 6\% of events. All
resolutions are shown after subtracting off the resolution of \ovts,
which is estimated from the \ov strip widths and the effects of multiple
scattering.

The resolution gradually worsens as the muon's path length
through the detector decreases. The addition of \ov information
becomes more important as fewer \pmts are used. The lowest energy
scintillator-dominated events are fit with a resolution of about
100(150)\,mm with(without) \ov hits. Muons that pass only through the
\iv and \buf\ --- \cher events, 35\% of the reconstructed muons --- are
fit with an $x$ and $y$ resolution of typically 200\,mm at the center
of their path without \ov hits, or 150\,mm when the \ov is used. The
29\% of muons passing only through the \iv are reconstructed with a
resolution of 250\,mm at the corner of the \iv that they intersect, but
very little angular information can be obtained without the \ov and so
the resolution is very poor at any other height. With \ov hits, the $x$
and $y$ resolution are better than 1\,m everywhere above the bottom of
the detector.

The presence of a shower distorts the fit by adding additional light
away from the muon track. However, if the shower begins inside the
detector, this light can only lead the muon's own light by a small
amount, and can sometimes be rejected by the \cher removal procedure.
Resolution for showering muons is typically 10--40\% worse than for
minimum ionizing muons, depending on the extent of the shower. The
most important application of muon tracking in \dc, \linine/\hee
identification, is affected by this loss of resolution since these
isotopes are produced by showering muons. However, the typical physical
distance between the muon and the \linine/\hee production is still
greater than the resolution, so the ability to identify them remains
excellent.

The accuracy of the stopping muon fit was tested by comparing the
reconstructed stopping position to the location of the following Michel
electron. The resolution was found to be 150\,mm in each of $x$, $y$
and $z$. Interestingly, while stopping muon fits done using \xyo from a
single \ov panel to constrain the track have somewhat better resolution
than those without \ov data, fits done with an \ovt have significantly
worse resolution. This is because the muon's path typically deviates
significantly from a straight line as it comes to a stop. While the
fit always assumes a straight-line path, if allowed to move it finds a
better approximation to the true path than that provided by the \ovt.

\section{Conclusions}\label{conclusions}

We have developed a sophisticated muon reconstruction algorithm for
\dc that provides resolution sufficient for several physics goals.
\dc's \ov was
instrumental in the development of this reconstruction and provides
substantial improvement to its resolution for tracks crossing
the \ov layers.
The techniques presented here are applicable, in whole or in part, to
any similar detector, such as those operated by the currently-running
Daya Bay~\cite{dayabay} and RENO~\cite{reno} experiments, or planned
detectors such as LENA~\cite{lena}, JUNO~\cite{juno},
RENO-50~\cite{reno50}, and SNO+~\cite{snoplus}.

\section{Acknowledgments}

We thank the French electricity company EDF; the European fund FEDER;
the R\'egion de Champagne Ardenne; the D\'epartement des Ardennes;
and the Communaut\'e des Communes Ardennes Rives acknowledge the
support of the CEA, CNRS/IN2P3, the computer center CCIN2P3, and
LabEx UnivEarthS in France (ANR-11-IDEX-0005-02); the Ministry of
Education, Culture, Sports, Science and Technology of Japan (MEXT) and
the Japan Society for the Promotion of Science (JSPS); the Department
of Energy and the National Science Foundation of the United States;
the Ministerio de Ciencia e Innovaci\'on (MICINN) of Spain; the Max
Planck Gesellschaft, and the Deutsche Forschungsgemeinschaft DFG (SBH
WI 2152), the Transregional Collaborative Research Center TR27, the
excellence cluster ``Origin and Structure of the Universe'', and the
Maier-Leibnitz-Laboratorium Garching in Germany; the Russian Academy
of Science, the Kurchatov Institute and RFBR (the Russian Foundation
for Basic Research); the Brazilian Ministry of Science, Technology and
Innovation (MCTI), the Financiadora de Estudos e Projetos (FINEP),
the Conselho Nacional de Desenvolvimento Cient\'ifico e Tecnol\'ogico
(CNPq), the S\~ao Paulo Research Foundation (FAPESP), and the Brazilian
Network for High Energy Physics (RENAFAE) in Brazil.

\bibliographystyle{model1-num-names.bst}

\bibliography{fido_nim}

\end{document}

%% file: authors.tex
\cortext[cor1]{Corresponding author. E-mail address: strait@hep.uchicago.edu (M.~Strait).}
\fntext[fn1]{Now at Department of Physics, University of Maryland,
College Park, Maryland 20742, USA.}
\fntext[fn2]{Now at Department of Physics, Kobe University, Kobe,
657-8501, Japan.}
\fntext[fn3]{Now at Department of Physics \& Astronomy, University of
Hawaii at Manoa, Honolulu, Hawaii 96822, USA.}
\fntext[fn4]{Now at Institut f\"{u}r Physik and Excellence Cluster
PRISMA, Johannes Gutenberg-Universit\"{a}t Mainz, 55128 Mainz, Germany.}

\address[Aachen]{III. Physikalisches Institut, RWTH Aachen 
University, 52056 Aachen, Germany}
\address[Alabama]{Department of Physics and Astronomy, University of 
Alabama, Tuscaloosa, Alabama 35487, USA}
\address[Argonne]{Argonne National Laboratory, Argonne, Illinois 
60439, USA}
\address[APC]{AstroParticule et Cosmologie, Universit\'{e} Paris 
Diderot, CNRS/IN2P3, CEA/IRFU, Observatoire de Paris, Sorbonne Paris 
Cit\'{e}, 75205 Paris Cedex 13, France}
\address[CBPF]{Centro Brasileiro de Pesquisas F\'{i}sicas, Rio de 
Janeiro, RJ, 22290-180, Brazil}
\address[Chicago]{The Enrico Fermi Institute, The University of 
Chicago, Chicago, Illinois 60637, USA}
\address[CIEMAT]{Centro de Investigaciones Energ\'{e}ticas, 
Medioambientales y Tecnol\'{o}gicas, CIEMAT, 28040, Madrid, Spain}
\address[Columbia]{Columbia University; New York, New York 10027, USA}
\address[Davis]{University of California, Davis, California 95616, USA}
\address[Drexel]{Department of Physics, Drexel University, Philadelphia,
Pennsylvania 19104, USA}
\address[Hiroshima]{Hiroshima Institute of Technology, Hiroshima, 
731-5193, Japan}
\address[IIT]{Department of Physics, Illinois Institute of 
Technology, Chicago, Illinois 60616, USA}
\address[INR]{Institute of Nuclear Research of the Russian Academy 
of Sciences, Moscow 117312, Russia}
\address[CEA]{Commissariat \`{a} l'Energie Atomique et aux Energies 
Alternatives, Centre de Saclay, IRFU, 91191 Gif-sur-Yvette, France}
\address[Kansas]{Department of Physics, Kansas State University, 
Manhattan, Kansas 66506, USA}
\address[Kobe]{Department of Physics, Kobe University, Kobe, 
657-8501, Japan}
\address[Kurchatov]{NRC Kurchatov Institute, Moscow 123182, Russia}
\address[MIT]{Massachusetts Institute of Technology, Cambridge,
Massachusetts 02139, USA}
\address[MaxPlanck]{Max-Planck-Institut f\"{u}r Kernphysik, 69117 
Heidelberg, Germany}
\address[Niigata]{Department of Physics, Niigata University, Niigata, 
950-2181, Japan}
\address[NotreDame]{University of Notre Dame, Notre Dame, Indiana 46556,
USA}
\address[IPHC]{IPHC, Universit\'{e} de Strasbourg, CNRS/IN2P3, 67037
Strasbourg, France}
\address[SUBATECH]{SUBATECH, CNRS/IN2P3, Universit\'{e} de Nantes, 
Ecole des Mines de Nantes, 44307 Nantes, France}
\address[Tennessee]{Department of Physics and Astronomy, University 
of Tennessee, Knoxville, Tennessee 37996, USA}
\address[TohokuUni]{Research Center for Neutrino Science, Tohoku 
University, Sendai 980-8578, Japan}
\address[TohokuGakuin]{Tohoku Gakuin University, Sendai, 981-3193, 
Japan}
\address[TokyoInst]{Department of Physics, Tokyo Institute of 
Technology, Tokyo, 152-8551, Japan  }
\address[TokyoMet]{Department of Physics, Tokyo Metropolitan 
University, Tokyo, 192-0397, Japan}
\address[Muenchen]{Physik Department, Technische Universit\"{a}t 
M\"{u}nchen, 85748 Garching, Germany}
\address[Tubingen]{Kepler Center for Astro and Particle Physics, 
Universit\"{a}t T\"{u}bingen, 72076 T\"{u}bingen, Germany}
\address[UFABC]{Universidade Federal do ABC, UFABC, Santo 
Andr\'{e}, SP, 09210-580, Brazil}
\address[UNICAMP]{Universidade Estadual de Campinas-UNICAMP, 
Campinas, SP, 13083-970, Brazil}
\address[vtech]{Center for Neutrino Physics, Virginia Tech, Blacksburg,
Virginia 24061, USA\\~} 

\author[TokyoInst]{Y.~Abe}
\author[CBPF]{J.C.~dos Anjos}
\author[CEA]{J.C.~Barriere}
\author[IPHC]{E.~Baussan}
\author[Aachen]{I.~Bekman}
\author[Davis]{M.~Bergevin}
\author[TohokuUni]{T.J.C.~Bezerra}
\author[INR]{L.~Bezrukov}
\author[Chicago]{E.~Blucher}
\author[MaxPlanck]{C.~Buck}
% ...
